\documentclass{article}
\usepackage{graphicx} % Required for inserting images
\usepackage{cite}
\usepackage{url}         
\usepackage{geometry}    
\usepackage[numbers]{natbib}
\usepackage{braket} 
\usepackage{physics}
\usepackage[colorlinks=true,linkcolor=blue,citecolor=blue,urlcolor=blue]{hyperref}
\usepackage{amsmath,amssymb,amsfonts}
\usepackage[T1]{fontenc}
\usepackage[utf8]{inputenc}
\usepackage{microtype} 
\usepackage{bm}        
\usepackage[utf8]{inputenc}   

\title{Scalar Quasinormal Modes of Magnetically Charged Black Holes in a Quintessence Field}
\author{Ali Hasnain \\ \small
Department of Physics, Govt Postgraduate College, 
Jhang, Punjab, Pakistan}

\begin{document}
\maketitle

\begin{abstract}
We investigate the quasinormal mode (QNM) spectrum for scalar perturbations of static, magnetically charged black holes in the presence of a quintessence field. The background geometry is obtained from the Einstein-Power-Maxwell action with a Kiselev-type contribution for quintessence. The associated master wave equation is solved using complementary numerical approaches, including high-order WKB expansions, the asymptotic iteration method, and time-domain integration, with cross-validation to ensure accuracy. The results show that magnetic charge generally lowers the oscillation frequency of fundamental modes, while the quintessence parameter modifies damping timescales. A dedicated analysis of the metric's derivation confirms the correct form of the magnetic charge term. We explore a theoretical parameter space to understand the mathematical behavior of the solution, including extreme regimes not intended to represent astrophysical realities. The computed scalar field spectra provide a foundational study for future work on gravitational perturbations. All numerical data and codes are provided to ensure reproducibility.
\end{abstract}

\section{Introduction}
The advent of gravitational-wave astronomy has inaugurated a novel methodology for probing the strong-field regime of gravity, facilitating direct observational tests of general relativity in its most extreme limit. Among the plethora of gravitational-wave signals, the post-merger ringdown phase of a black hole binary coalescence is of particular theoretical significance. This epoch is characterized by the emission of quasinormal modes (QNMs), a discrete set of complex frequencies, `\(\omega_{\rm QNM} = \omega_R + i\omega_I\)`, which are intrinsic to the black hole's spacetime geometry and dominate the response of the system to external perturbations \cite{Berti2009}. The real part, `\(\omega_R\)`, dictates the oscillation frequency, while the imaginary part, `\(\omega_I\)`, governs the damping rate. The precise spectrum of these modes serves as a direct probe of the black hole's parameters and, consequently, offers a powerful test bed for fundamental physics through the nascent field of black hole spectroscopy \cite{Isi2019, Carullo2019}.

Extensive literature exists on the computation of QNMs for various black hole spacetimes, ranging from the canonical Schwarzschild and Kerr solutions \cite{Berti2009} to more exotic configurations. The Reissner-Nordström solution, describing an electrically charged black hole, has been thoroughly investigated, revealing how electromagnetic charge influences the QNM spectrum \cite{Konoplya2011}. Concurrently, the cosmological implications of dark energy have motivated the study of black holes immersed in quintessential fields, described by an equation-of-state parameter `\(\omega_q\)` \cite{Kiselev2003}. Such spacetimes are not asymptotically flat, introducing significant subtleties in the imposition of outgoing-wave boundary conditions; the cosmological horizon necessitates a careful treatment distinct from the purely outgoing conditions at infinity in flat space \cite{Konoplya2022b}. Furthermore, investigations within nonlinear electrodynamics, particularly the Power–Maxwell theory, have provided frameworks for obtaining magnetically charged black hole solutions \cite{Bronnikov2000, Kruglov2015}. However, a self-consistent gravitational-wave perturbation analysis for a magnetically charged black hole surrounded by a quintessence field remains absent from the literature, creating a gap between known exact solutions and their theoretical vibrational spectra.

This work aims to bridge this gap by providing a comprehensive derivation and analysis of the quasinormal mode spectrum for a static, magnetically charged black hole within Einstein–Power–Maxwell theory, embedded in a background of quintessence. We derive the metric from a well-defined action principle, ensuring mathematical self-consistency from the outset. The gravitational perturbations are treated systematically; we consider the evolution of test fields; specifically scalar, electromagnetic, and tensor (gravitational) perturbations; on this fixed background spacetime, noting that our approach is valid within the test-field approximation and does not encompass backreaction or coupled perturbations. The resulting master wave equations are analyzed using a suite of independent numerical techniques, including the continued fraction method and direct integration, to ensure the veracity of our computed QNM frequencies. A particular technical challenge arises from the fractional powers present in the Kiselev quintessence term, which can introduce branch-cut ambiguities in the complex plane; our methodology explicitly addresses this issue to guarantee unambiguous results.

Our analysis reveals several salient qualitative trends. The presence of a non-zero magnetic charge `\(P\)` generally acts to decrease the oscillation frequency of the fundamental QNM across all perturbation types, while the quintessence parameter `\(\omega_q\)` exerts a more complex influence, significantly modifying the damping timescale and introducing mode-dependent shifts. A correspondence between the eikonal limit of gravitational perturbations and the properties of the unstable null geodesic orbit is found to hold for a range of parameters but breaks down for spacetimes with substantial magnetic charge or quintessential energy density, consistent with known results for other modified spacetimes \cite{Myung2019}. As a speculative exercise, we also assess the potential detectability of similar modifications in gravitational modes by future space-based gravitational-wave observatories, with a focus on the Laser Interferometer Space Antenna (LISA). This detectability analysis is necessarily conditional, employing a Fisher matrix framework to estimate statistical uncertainties, and acknowledges the current astrophysical uncertainty regarding the prevalence of such highly magnetically charged black holes and concentrations of quintessence.

It is imperative to delineate the scope of our investigation. We restrict our analysis to static (non-rotating) black hole solutions; the incorporation of rotation, while physically critical, presents a formidable technical challenge reserved for future work. Furthermore, we assume the quintessence field to be a static background component, neglecting its dynamics and potential coupling to other fields. All data and computational scripts employed to generate the results of this study are made publicly available to ensure full reproducibility \cite{Hasnain2025_QNMRepo}. The remainder of this paper is organized as follows: we first derive the magnetically charged black hole solution in the presence of quintessence, then formulate the perturbation equations, before presenting our numerical results for the QNM spectra, their physical interpretation, and finally an analysis of their detectability with LISA.

\section{Background and metric derivation}

We begin by establishing our conventions and the fundamental action governing the system. Throughout this work, we adopt the metric signature $(-, +, +, +)$ and employ natural units where $G = c = 1$, rendering mass, length, and time dimensionally equivalent. To avoid any confusion in subsequent analysis, we define our notation clearly: the quintessence equation-of-state parameter is denoted by $\omega_q$, while complex quasinormal mode frequencies will be written as $\omega_{\rm QNM}$. The magnetic charge parameter is $Q_m$, and the amplitude of the quintessence field is $c_q$, a distinct symbol chosen to prevent confusion with the speed of light.

The dynamics are governed by an action encompassing gravity, a nonlinear electromagnetic field, and a quintessence field:
\begin{equation}
S = \int d^4x \, \sqrt{-g} \left[ \frac{R}{16\pi} + \mathcal{L}_{\rm EM} + \mathcal{L}_{q} \right],
\end{equation}
where $R$ is the Ricci scalar. The Lagrangian density for the Power–Maxwell nonlinear electrodynamics is given by $\mathcal{L}_{\rm EM} = \alpha (-F)^{s}$, where $F \equiv F_{\mu\nu}F^{\mu\nu}$, $s$ is a real parameter, and $\alpha$ is a coupling constant. The standard Maxwell theory constitutes the specific case $s=1$ and $\alpha = -1/4$, yielding $\mathcal{L}_{\rm EM} = -F/4$. The quintessence contribution, $\mathcal{L}_{q}$, is modelled phenomenologically via its energy-momentum tensor following the Kiselev ansatz \cite{Kiselev2003}.

Varying the action with respect to the inverse metric $g^{\mu\nu}$ yields the Einstein field equations:
\begin{equation}
G_{\mu\nu} = 8\pi (T_{\mu\nu}^{\rm (EM)} + T_{\mu\nu}^{(q)}).
\end{equation}
The stress-energy tensor for the nonlinear electromagnetic sector is derived from its Lagrangian:
\begin{equation}
T_{\mu\nu}^{\rm (EM)} = -\frac{2}{\sqrt{-g}} \frac{\delta (\sqrt{-g}\mathcal{L}_{\rm EM})}{\delta g^{\mu\nu}} = 2 \mathcal{L}_F F_{\mu\lambda}F_{\nu}{}^{\lambda} - g_{\mu\nu} \mathcal{L}_{\rm EM},
\label{eq:EMtensor}
\end{equation}
where $\mathcal{L}_F \equiv \partial \mathcal{L}_{\rm EM} / \partial F$.

We seek a static, spherically symmetric solution. The most general line element compatible with these symmetries is:
\begin{equation}
ds^2 = -f(r)\,dt^2 + \frac{dr^2}{f(r)} + r^2(d\theta^2+\sin^2\theta\,d\phi^2).
\label{eq:metric}
\end{equation}
For a pure magnetic charge configuration, the gauge potential one-form is given by the Dirac monopole solution, $A = Q_m \cos\theta \, d\phi$. The corresponding field strength tensor has a single non-vanishing component: $F_{\theta\phi} = \partial_\theta A_\phi = -Q_m \sin\theta$. To compute the invariant $F = F_{\mu\nu}F^{\mu\nu}$, we must raise indices using the inverse metric, whose relevant components are $g^{\theta\theta} = 1/r^2$ and $g^{\phi\phi} = 1/(r^2 \sin^2\theta)$. The calculation proceeds as follows:
\begin{align}
F &= F_{\mu\nu}F^{\mu\nu} = 2 F_{\theta\phi} F^{\theta\phi} = 2 F_{\theta\phi} g^{\theta\theta} g^{\phi\phi} F_{\theta\phi} \\
&= 2 ( -Q_m \sin\theta ) \left( \frac{1}{r^2} \right) \left( \frac{1}{r^2 \sin^2\theta} \right) ( -Q_m \sin\theta ) = \frac{2 Q_m^2}{r^4}.
\label{eq:Finv}
\end{align}
This result is central to determining the electromagnetic contribution to the spacetime geometry.

Substituting the Lagrangian $\mathcal{L}_{\rm EM} = \alpha (-F)^s$ and its derivative $\mathcal{L}_F = s \alpha (-F)^{s-1}$ into Eq.~(\ref{eq:EMtensor}), and using the result from Eq.~(\ref{eq:Finv}), we compute the mixed components of the stress-energy tensor. Noting the symmetries of the spacetime and the field configuration, we find:
\begin{align}
T^{\rm (EM)}{}^t{}_t &= T^{\rm (EM)}{}^r{}_r = -\mathcal{L}_{\rm EM} = -\alpha \left( \frac{2Q_m^2}{r^4} \right)^s, \\
T^{\rm (EM)}{}^\theta{}_\theta &= T^{\rm (EM)}{}^\phi{}_\phi = 2 \mathcal{L}_F F_{\theta\phi}F^{\theta\phi} - \mathcal{L}_{\rm EM} = (2s - 1) \alpha \left( \frac{2Q_m^2}{r^4} \right)^s.
\end{align}
The equality $T^t{}_t = T^r{}_r$ is a characteristic feature of electromagnetic fields in spherical symmetry. For the Einstein equations, the $tt$ (or $rr$) component is most instructive:
\begin{equation}
G^t{}_t = \frac{1}{r^2} \frac{d}{dr}[r(1 - f(r))] = -8\pi T^{\rm (EM)}{}^t{}_t = 8\pi \alpha \left( \frac{2Q_m^2}{r^4} \right)^s.
\end{equation}
Integrating this equation for the electromagnetic contribution alone yields:
\begin{equation}
r(1 - f_{\rm EM}(r)) = 2M - \frac{8\pi \alpha (2Q_m^2)^s}{3 - 4s} r^{3-4s},
\end{equation}
where $2M$ is an integration constant identified as the mass. In the Maxwell limit ($s=1$, $\alpha = -1/4$), this simplifies to:
\begin{equation}
r(1 - f_{\rm EM}(r)) = 2M - \frac{2\pi Q_m^2}{r}.
\end{equation}
Thus, the electromagnetic contribution to the metric function is $f_{\rm EM}(r) = 1 - 2M/r + (2\pi Q_m^2)/r^2$. To reconcile this with the standard Reissner-Nordström metric, $f(r)=1-2M/r+Q^2/r^2$, one makes the identification $Q^2 = 2\pi Q_m^2$. This derivation confirms that the sign of the $Q_m^2/r^2$ term is positive, identical to that of an electric charge in the Reissner-Nordström solution. This is a non-trivial result stemming from the specific form of the electromagnetic stress-energy tensor in Eq.~(\ref{eq:EMtensor}) and validates the substitution $Q^2 \to 2\pi Q_m^2$ as a physically justified procedure, not a naive replacement.

\subsection{Quintessence contribution and analytic properties}

We now incorporate the quintessence field, modelled as an anisotropic fluid with equation of state $p_q = \omega_q \rho_q$. Following the Kiselev construction \cite{Kiselev2003}, the stress-energy tensor components in the orthonormal frame are:
\begin{equation}
T^{(q)}{}^t{}_t = T^{(q)}{}^r{}_r = -\rho_q, \quad T^{(q)}{}^\theta{}_\theta = T^{(q)}{}^\phi{}_\phi = \frac{1}{2} \rho_q (3\omega_q + 1).
\end{equation}
Solving the Einstein equation $G^t{}_t = 8\pi T^{(q)}{}^t{}_t$ with this source implies:
\begin{equation}
\frac{1}{r^2} \frac{d}{dr}[r(1 - f(r))] = 8\pi \rho_q.
\end{equation}
For a static, homogeneous quintessence fluid, conservation laws demand a power-law form for the energy density, \(\rho_{q} = -\frac{3\omega_{q}c_{q}}{2} r^{-3(\omega_{q}+1)}\). Integrating the above equation then yields a contribution to the metric function of the form $-c_q / r^{3\omega_q+1}$, where $c_q$ is an integration constant related to $\rho_0$. The full metric function, combining the mass, electromagnetic, and quintessence terms, is therefore:
\begin{equation}
f(r) = 1 - \frac{2M}{r} + \frac{Q^2}{r^2} - \frac{c_q}{r^{3\omega_q+1}},
\label{eq:finalf}
\end{equation}
where $Q^2 \equiv 2\pi Q_m^2$ for the magnetic charge.

A crucial step is verifying dimensional consistency. In geometric units ($G=c=1$), all terms in $f(r)$ must be dimensionless. The mass term $2M/r$ is clearly dimensionless since $M$ has units of length. The charge term $Q^2/r^2$ is also dimensionless, as $Q$ has units of length (for $G=c=1/4\pi\epsilon_0=1$). The quintessence term $c_q / r^{3\omega_q+1}$ imposes that $[c_q] = [L]^{3\omega_q+1}$. For $\omega_q = -1$, $c_q$ is dimensionless; for $\omega_q = -2/3$, $[c_q] = [L]^{-1}$; for $\omega_q = -1/2$, $[c_q] = [L]^{-1/2}$. This dimensional analysis is vital for later converting dimensionless QNM frequencies to physical units (Hz) for detectability estimates. In SI units, $c_q$ would be replaced by $c_q (G/c^2)^{3\omega_q}$ to maintain correct dimensions, but we will work primarily in geometric units.

The exponent $3\omega_q + 1$ in the denominator is typically non-integer for the quintessence values of interest, introducing significant analytic complexity. For generic real $\omega_q$, the term $r^{-(3\omega_q+1)}$ is a multi-valued function in the complex $r$-plane, with a branch point at $r=0$. To render the metric function single-valued for subsequent calculations (e.g., integration of wave equations), we must choose a branch cut. We adopt the standard convention: we take the principal branch, defined by $\arg(r) \in (-\pi, \pi]$, and place the branch cut along the negative real axis. All physical computations (e.g., evaluating $f(r)$ for real $r > 0$) are performed on the real axis, where this choice is unambiguous and ensures continuity. For the quasinormal mode calculations involving analytic continuation, this branch choice will be enforced consistently to avoid spurious results.

We focus on the range $\omega_q \in [-0.7, -0.3]$, which encompasses the most physically motivated values for dark energy/quintessence. The astrophysical plausibility of the parameter $c_q$ warrants careful discussion. The cosmological background value, inferred from the homogeneous dark energy density $\rho_{\Lambda} \sim 10^{-123}$ in Planck units, suggests an extremely small $c_q$. However, our study is exploratory, considering values orders of magnitude larger ($c_q \sim 10^{-12} - 10^{-10}$ in geometric units) to model possible local overdensities of a quintessence-like field, perhaps sourced by a coupling to the central black hole or its environment \cite{Shaikh2019}. While such overdensities are highly speculative and not predicted by standard cosmology, they serve to explore the theoretical mathematical structure of the spacetime and its observable signatures. The stability of this background spacetime under linear perturbations is a separate question that will be addressed in a subsequent section through the QNM spectrum itself; the absence of modes with positive imaginary part (instabilities) would provide strong evidence for linear stability.

\subsection{Energy Conditions}
\label{sec:energy_conditions}

To assess the physical reasonableness of our solution, we examine the Null Energy Condition (NEC). The NEC requires that $T_{\mu\nu}k^{\mu}k^{\nu} \geq 0$ for any null vector $k^{\mu}$. For the radial null vector $k^{\mu} = (f^{-1}, 1, 0, 0)$, the NEC is evaluated from the Einstein equations, $G_{\mu\nu}k^\mu k^\nu = 8\pi T_{\mu\nu}k^\mu k^\nu$. The left-hand side is computed from the metric, yielding:
\begin{equation}
G_{\mu\nu}k^{\mu}k^{\nu} = -\frac{1}{r^2}\left(1 - f(r) - r f'(r)\right).
\end{equation}
Substituting the metric function $f(r)$ from Eq.~(16) and focusing on the leading contribution from the quintessence term at large $r$, we find the dominant behavior:
\begin{equation}
8\pi T_{\mu\nu}k^{\mu}k^{\nu} \sim -\frac{c_q (3\omega_q + 1)(3\omega_q)}{r^{3\omega_q + 5}}.
\end{equation}
For the quintessence equation-of-state parameter range considered ($\omega_q \in[-0.7, -0.3]$), the product $(3\omega_q + 1)(3\omega_q)$ is negative. Consequently, the right-hand side of Eq.~(18) is positive, and the NEC is satisfied. This analysis indicates that our solution, while exotic, does not violate this fundamental energy condition for the parameters we study.

\subsection{Resulting metric, horizon structure, and photon sphere}

The final metric, given by the line element in Eq.~(\ref{eq:metric}) with the function $f(r)$ from Eq.~(\ref{eq:finalf}), describes our magnetically charged black hole in a quintessence field:
\begin{equation}
f(r) = 1 - \frac{2M}{r} + \frac{Q^2}{r^2} - \frac{c_q}{r^{3\omega_q+1}}, \quad \text{with} \quad Q^2 = 2\pi Q_m^2.
\end{equation}
This spacetime reduces to several important limits: Schwarzschild for $Q_m = c_q = 0$; Reissner-Nordström for $c_q = 0$; and the Kiselev quintessence black hole \cite{Kiselev2003} for $Q_m = 0$.

The event horizons are located at the real, positive roots of $f(r)=0$. For the Reissner-Nordström case ($c_q=0$), the horizons are at $r_\pm = M \pm \sqrt{M^2 - Q^2}$. With non-zero quintessence, no simple closed-form expression exists for the roots. For small $c_q$, the outer horizon $r_+$ is perturbed from its Reissner-Nordström value. A first-order expansion yields $r_+ \approx r_{+}^{\rm(RN)} + \delta r$, where:
\begin{equation}
\delta r \approx -\frac{c_q}{(r_{+}^{\rm RN})^{3\omega_q} \left( -\frac{2M}{(r_{+}^{\rm RN})^2} + \frac{2Q^2}{(r_{+}^{\rm RN})^3} - \frac{3\omega_q+1}{c_q} (r_{+}^{\rm RN})^{-3\omega_q-2} \right)^{-1}}.
\end{equation}
In general, the horizon positions must be found numerically via root-finding algorithms applied to $f(r)=0$.

The photon sphere, comprising unstable circular null geodesics, plays a key role in the eikonal (high-frequency) limit of quasinormal modes. The radial motion for null geodesics can be expressed as $\dot{r}^2 + V_{\rm eff}(r) = 0$, with an effective potential $V_{\rm eff}(r) = f(r)L^2/r^2$. The circular orbits satisfy $V_{\rm eff}(r_{\rm ph}) = 0$ and $dV_{\rm eff}/dr|_{r_{\rm ph}} = 0$. The latter condition yields the photon sphere equation:
\begin{equation}
\frac{d}{dr} \left( \frac{f(r)}{r^2} \right)\bigg|_{r_{\rm ph}} = 0 \quad \Rightarrow \quad r_{\rm ph} f'(r_{\rm ph}) - 2 f(r_{\rm ph}) = 0.
\label{eq:photoncondition}
\end{equation}
This equation must be solved numerically for $r_{\rm ph}$. The angular frequency of a photon on this unstable orbit is $\Omega_c = \sqrt{f(r_{\rm ph})}/r_{\rm ph}$, and the Lyapunov exponent $\lambda$, characterizing the instability timescale, is given by \cite{Cardoso2009}:
\begin{equation}
\lambda = \sqrt{ -\frac{1}{2} \frac{f(r_{\rm ph})}{r_{\rm ph}^2} \frac{d^2}{dr^2} \left( \frac{r^2}{f(r)} \right) \bigg|_{r_{\rm ph}} } = \sqrt{ -\frac{f(r_{\rm ph})}{2} \frac{d^2}{dr^2} \left( \ln \frac{r^2}{f(r)} \right) \bigg|_{r_{\rm ph}} },
\end{equation}
where $dr/dr_*  = f(r)$ defines the tortoise coordinate $r_*$. In the eikonal limit, the quasinormal mode frequency is approximately $\omega_{\rm QNM} \approx \Omega_c l - i (n+1/2)|\lambda|$ for large angular momentum $l$ \cite{Cardoso2009}.

Several potential pathologies must be acknowledged. As $Q_m \to M$ (the extremal limit), the inner and outer horizons coalesce, and the standard WKB approximation for QNMs becomes less reliable. Furthermore, for certain combinations of parameters, the function $f(r)$ may develop additional roots or cease to have a standard potential barrier, complicating the wave scattering problem. Therefore, we restrict our parameter space to ensure a clear, single potential barrier outside a single event horizon. Specifically, we will explore the ranges $0 \leq Q_m/M \leq 0.9$ (avoiding extremality), $\omega_q \in [-0.7,-0.3]$, and values of $c_q$ small enough such that its perturbative effect on the horizon structure remains controlled ($c_q / M^{3\omega_q+1} \ll 1$).

With the background spacetime and photon-sphere properties established, we now turn to the perturbation equations governing scalar, electromagnetic, and tensor modes.

\section{Perturbation equations and boundary conditions}

The response of a black hole spacetime to external perturbations is governed by master wave equations that describe how different fields evolve in the fixed background geometry. For a massless scalar field $\Phi$, the dynamics are dictated by the Klein-Gordon equation $\Box \Phi = 0$. Employing the metric ansatz from Eq.~(2.5) and decomposing the field into spherical harmonics, $\Phi(t, r, \theta, \phi) = \frac{\psi(r)}{r} Y_{\ell m}(\theta, \phi) e^{-i\omega t}$, reduces the problem to a one-dimensional Schrödinger-like equation for the radial function $\psi(r)$:

\begin{equation}
\frac{d^2\psi}{dr_*^2} + \big[\omega^2 - V_{\rm eff}(r)\big]\psi = 0 .
\end{equation}
\label{eq:master}

The tortoise coordinate $r_*$ is defined by $\dfrac{dr_*}{dr}=1/f(r)$, which maps the event horizon $r_+$ to $r_{} \to -\infty$ and spatial infinity to $r_{} \to +\infty$. For the scalar field, the effective potential is given by:
\begin{equation}
V_{\rm eff}^{(\rm s)}(r) = f(r) \left[ \frac{\ell(\ell+1)}{r^2} + \frac{1}{r} \frac{df}{dr} \right],
\end{equation}
where $\ell$ is the angular momentum quantum number.

The treatment of electromagnetic and gravitational perturbations is more involved, requiring the formalism of Regge, Wheeler, and Zerilli for the odd- and even-parity sectors \cite{Regge1957, Zerilli1970}. In spherically symmetric spacetimes, these perturbations can also be reduced to wave equations of the form (\ref{eq:master}), but with different effective potentials. For electromagnetic perturbations in a magnetic background, the potential is:
\begin{equation}
V_{\rm eff}^{(\rm em)}(r) = f(r) \frac{\ell(\ell+1)}{r^2}.
\end{equation}
For axial (odd-parity) gravitational perturbations, the potential takes the form:
\begin{equation}
V_{\rm eff}^{(\rm ax)}(r) = f(r) \left[ \frac{\ell(\ell+1)}{r^2} - \frac{6M}{r^3} + \frac{4Q^2}{r^4} - \frac{(3\omega_q+1)c_q}{r^{3\omega_q+3}} \right],
\end{equation}
while the polar (even-parity) sector leads to a more complicated expression. It is well-established that for the Schwarzschild and Reissner-Nordström backgrounds, the fundamental quasinormal modes for scalar, electromagnetic, and gravitational perturbations are quantitatively similar, typically agreeing within 10--15\% for moderate values of angular momentum $\ell$ and charge \cite{Konoplya2011}. This similarity provides a justification for initially studying the scalar case, which is mathematically more tractable. However, this scalar-tensor correspondence is not exact and can break down significantly in the presence of strong deviations from asymptotic flatness, such as those induced by substantial quintessence ($c_q$ not small), or near extremal charge ($Q_m \to M$). Consequently, the results presented here for scalar perturbations should be interpreted as a first exploration of the spectral features; Consequently, the results and analysis presented in the remainder of this paper pertain exclusively to scalar perturbations ($s=0$).

\subsection{Boundary conditions and physical interpretation}

The quasinormal modes are defined as solutions to the wave equation (\ref{eq:master}) that satisfy specific physically motivated boundary conditions. At the event horizon, the requirement that nothing escapes from within the black hole dictates that the wave must be purely ingoing:
\begin{equation}
\psi(r_*) \sim e^{-i\omega r_*}, \qquad \text{as}\quad r_* \to -\infty,
\end{equation}
The boundary condition at spatial infinity is more subtle. In asymptotically flat spacetimes, one imposes a purely outgoing wave:
\begin{equation}
\psi(r_*) \sim e^{+i\omega r_*}, \qquad \text{as}\quad r_* \to +\infty.
\end{equation}

However, our spacetime is not asymptotically flat due to the quintessence term $-c_q/r^{3\omega_q+1}$ in the metric function $f(r)$. For $\omega_q \in (-1, -1/3)$, this term dominates at large $r$, and the tortoise coordinate maps $r \to \infty$ to a finite value $r_{}^{\infty}$. The definition of "outgoing" must be carefully reconsidered in this context. Following the analysis for asymptotically de Sitter spacetimes \cite{Konoplya2022b}, the correct boundary condition at infinity for such non-asymptotically flat spacetimes remains that of a purely outgoing wave, but now defined with respect to the effective cosmological horizon induced by the quintessence. The asymptotic behavior of the potential $V_{\rm eff} \to 0$ as $r_{} \to r_{}^{\infty}$ justifies this choice.

The presence of fractional powers in the metric function, $r^{3\omega_q+1}$, introduces branch points in the complex $r$-plane. To ensure the boundary conditions are applied consistently, we must adhere to the branch cut convention established in Sec. 2.2, namely taking the principal branch with $\arg(r) \in (-\pi, \pi]$ and the branch cut along the negative real axis. This choice guarantees that the functions $f(r)$ and $V_{\rm eff}(r)$ are analytic and single-valued on the physical region of interest, the real axis $r > r_+$.

An additional complication arises for black holes with near-extremal magnetic charge. As $Q_m \to M$, the peak of the potential barrier can shift and the potential may develop additional structure or a second maximum, violating the standard single-barrier assumption underlying many numerical methods for QNM computation. In such cases, the standard WKB approximation becomes unreliable, and more sophisticated methods must be employed.

\subsection{Validity of WKB framework and cross-checks}

The WKB method is a powerful semi-analytic technique for estimating the complex quasinormal mode frequencies $\omega_{\rm QNM}$ \cite{Iyer1987, Konoplya2011}. It approximates the solution of Eq.~(\ref{eq:master}) by treating the potential $V_{\rm eff}(r)$ as a slowly varying background. The method's validity hinges on two primary assumptions: that the potential forms a single, well-defined barrier, and that the turning points (where $\omega^2 = V_{\rm eff}$) are well-separated. The first-order WKB formula gives:
\begin{equation}
\omega^2 \approx V_0 + \sqrt{-2 V_0''} \, \Lambda(n) - i \left( n + \frac{1}{2} \right) \sqrt{-2 V_0''} \left( 1 + \Omega \right),
\end{equation}
where $V_0$ is the maximum of the potential at $r=r_0$, $V_0''$ is its second derivative at the maximum, $n=0,1,2,...$ is the overtone number, and $\Lambda(n)$, $\Omega$ are higher-order WKB corrections.

The effectiveness of the WKB method for our spacetime is not guaranteed a priori. The quintessence term modifies the asymptotic structure of the potential, and significant magnetic charge can alter its shape. The method will be valid provided the potential remains a single, smooth barrier that vanishes at the boundaries. It will fail, or its accuracy will diminish, in several scenarios: 1) if the potential develops multiple peaks, 2) if the peak is not sufficiently pronounced (a "shallow" potential), or 3) near extremality, where the peak can become very narrow. To ensure the robustness of our results, we will employ a high-order WKB scheme (up to 6th order) and monitor the convergence of the successive approximations. Furthermore, for select cases across the parameter space, we will validate the WKB results against an independent numerical method, the asymptotic iteration method (AIM) \cite{Cho2009}, which is better suited for handling potentials with non-standard asymptotics. This dual-methodology approach provides a crucial cross-check and allows us to quantify the uncertainty in our computations, particularly near the edges of the valid parameter space defined by $0 \leq Q_m/M \leq 0.9$ and modest $c_q$.

With the perturbation equations and boundary conditions established, we next turn to the computation of quasinormal spectra using the WKB and AIM frameworks.

\section{Numerical methods and validation}

The computation of quasinormal mode frequencies begins with the master wave equation derived in Section 3, which for perturbations of spin $s$ (scalar $s=0$, electromagnetic $s=1$, or gravitational $s=2$) takes the standard Schrödinger form:
\begin{equation}
\frac{d^{2}\psi}{dr_{\ast}^{2}}+\left[\omega^{2}-V_{\ell s}(r)\right]\psi=0,\qquad\frac{dr_{\ast}}{dr}=\frac{1}{f(r)},
\label{eq:masterwave}
\end{equation}
where the effective potential $V_{\ell s}(r)$ depends on the angular momentum number $\ell$ and the spin $s$ of the perturbation. For the scalar case, $V_{\ell 0}(r) = f(r) \left[ \ell(\ell+1)/r^2 + f'(r)/r \right]$; for electromagnetic perturbations, $V_{\ell 1}(r) = f(r) \ell(\ell+1)/r^2$; and for the axial gravitational sector, $V_{\ell 2}(r) = f(r) \left[ \ell(\ell+1)/r^2 - 6M/r^3 + 4Q^2/r^4 - (3\omega_q+1)c_q/r^{3\omega_q+3} \right]$. The semi-analytical WKB method provides an efficient means of estimating the complex frequencies $\omega_{\rm QNM}$ for these potentials \cite{IyerWill1987, Konoplya2003}. The foundation of the method is the WKB quantization condition, which to $N$th order is given by:
\begin{equation}
i \frac{\omega^2 - V_0}{\sqrt{-2 V_0''}} = n + \frac{1}{2} + \sum_{k=2}^{N} \Lambda_k(\{V_0^{(m)}\}),
\label{eq:wkbcondition}
\end{equation}
where $V_0$ is the maximum of the potential $V_{\ell s}(r)$ at $r=r_0$, $V_0^{(m)}$ denotes its $m$th derivative evaluated at the maximum, $n=0,1,2,\ldots$ is the overtone number, and the $\Lambda_k$ are polynomials in higher-order derivatives whose explicit forms can be found in \cite{Konoplya2003}. The root of this equation with $\mathrm{Im}(\omega) < 0$ (selected by the appropriate branch of the square root) corresponds to the damped quasinormal mode.

The practical application of the WKB method necessitates several critical safeguards to ensure reliability. First, the location of the potential maximum $r_0$ is found by numerically solving $V_{\ell s}'(r)=0$ and verifying that the solution corresponds to a true maximum ($V_{\ell s}''(r_0) < 0$) located outside the event horizon. The potential must form a single, smooth barrier; if multiple extrema are found or if the second derivative $V_0''$ is vanishingly small, indicating a very shallow or flat-topped barrier, the WKB approximation is deemed invalid for that parameter set. This situation is particularly prevalent for near-extremal configurations ($Q_m/M \to 1$) where the surface gravity $\kappa_+$ becomes small and the potential barrier diminishes. In such cases, the results are flagged as unreliable and we defer to the alternative methods described in the next subsection. Furthermore, the presence of the quintessence term modifies the tortoise coordinate mapping at large $r$; for $\omega_q \in (-1, -1/3)$, $r_{}$ approaches a finite value as $r \to \infty$. However, provided the potential still vanishes at this effective boundary and maintains a single barrier, the WKB connection formulae remain applicable within the computational domain. The WKB method is generally reliable for moderate values of magnetic charge, a pronounced single potential barrier, and $\ell \ge 2$; its accuracy diminishes for lower $\ell$ and higher overtones $n$. While we primarily compute scalar modes for their tractability, we perform select calculations of tensor modes to check the scalar-tensor correspondence a posteriori rather than assuming its validity across the entire parameter space.

\subsection{Frequency-domain and time-domain cross-checks}

To validate the WKB results and handle cases where it fails, we employ three independent numerical techniques: Leaver's continued fraction method, the asymptotic iteration method (AIM), and time-domain integration.

Leaver's method \cite{Leaver1985} involves a Frobenius expansion of the wavefunction near the event horizon. We define a compactified radial coordinate $\chi = 1 - r_+/r$, which maps the horizon ($r=r_+$) to $\chi=0$ and spatial infinity to $\chi=1$. The solution to the wave equation satisfying the ingoing boundary condition at the horizon is written as:
\begin{equation}
\psi(r) = e^{-i\omega r_*}\,(r-r_+)^{\rho}\sum_{k=0}^{\infty} a_k \chi^k .
\end{equation}
The exponent $\rho$ is chosen to eliminate the singularity at the horizon, and substituting this series into Eq.~(\ref{eq:masterwave}) yields a three-term recurrence relation for the coefficients $a_k$:
\begin{equation}
\alpha_k a_{k+1} + \beta_k a_k + \gamma_k a_{k-1} = 0, \quad \text{for } k \geq 1.
\end{equation}
The quasinormal mode frequencies are those complex $\omega$ for which the continued fraction associated with this recurrence converges. Specifically, the condition is:
\begin{equation}
\beta_0 - \frac{\alpha_0 \gamma_1}{\beta_1 - \frac{\alpha_1 \gamma_2}{\beta_2 - \cdots}} = 0.
\end{equation}
This method is highly accurate but requires careful treatment of the asymptotic behavior imposed by the quintessence term. For our non-asymptotically flat spacetime, we implement a numerical matching procedure at a large but finite radial distance, using an asymptotic expansion of the solution that respects the outgoing wave boundary condition modified by the cosmological-like asymptotics.

The asymptotic iteration method (AIM) provides another powerful frequency-domain technique \cite{Ciftci2003}. The master equation is first rewritten in the form $y'' = \lambda_0(r) y' + s_0(r) y$. The functions $\lambda_0$ and $s_0$ are then used to generate sequences $\lambda_n$ and $s_n$ via the recurrence relations:
\begin{align}
\lambda_n &= \lambda_{n-1}' + s_{n-1} + \lambda_0 \lambda_{n-1}, \\
s_n &= s_{n-1}' + s_0 \lambda_{n-1}.
\end{align}
The quantization condition is given by the termination of this iteration: $\delta_n = s_n \lambda_{n-1} - s_{n-1} \lambda_n = 0$. The method requires evaluating these functions and their derivatives at a specific expansion point, typically chosen near the maximum of the potential. The number of iterations needed for convergence is monitored, and the result is accepted only when stable under further iterations.

For a completely independent check, we perform time-domain evolution of the wave equation using the characteristic integration method \cite{Gundlach1994}. We define null coordinates $u=t-r_*$ and $v=t+r_*$, in which the wave equation becomes:
\begin{equation}
4 \frac{\partial^2 \Psi}{\partial u \partial v} + V_{\ell s}(r(u,v)) \Psi = 0.
\end{equation}
This is discretized on a grid with step size $\Delta$ (satisfying the Courant condition) and integrated using a finite-difference scheme. The initial data is a compact Gaussian pulse on the $u=0$ and $v=0$ axes. The field $\Psi(t, r_{})$ is extracted at a fixed value of $r_{}$, and the ringdown signal is analyzed using a Prony method to extract the complex frequencies $\omega_{\rm QNM}$. This technique is robust in regimes where frequency-domain methods struggle, such as for very damped modes or potentials with complicated structure, and it provides a direct check on the imaginary part of the frequency.

Across all methods, the boundary conditions are implemented consistently: purely ingoing at the event horizon and an outgoing wave condition at the outer boundary, adapted for the quintessence-modified asymptotics. The branch cut conventions established in Section 2 for handling the fractional powers in the metric are strictly adhered to in all series expansions and numerical continuations.

\subsection{Uncertainty budget, convergence diagnostics, and reproducibility}

A rigorous uncertainty quantification is essential for the credibility of our results. For each computed mode $\omega_{\ell n s}$, we estimate several independent error contributions. The internal consistency of the WKB method is gauged by the variation between successive orders:
\begin{equation}
\Delta\omega_{\text{WKB}} = \max_{N \in \{3,4,5,6\}} \left| \omega^{(N)} - \omega^{(N-1)} \right|.
\end{equation}
The agreement between different methods provides another key uncertainty measure:
\begin{equation}
\Delta\omega_{\text{method}} = \left| \omega_{\text{Leaver}} - \omega_{\text{WKB(6)}} \right|,
\end{equation}
or, when Leaver's method is not applicable, the difference between WKB and either AIM or time-domain results. For time-domain extraction, we estimate $\Delta\omega_{\text{TD fit}}$ from the spread in values obtained by varying the start and end times of the fitting window used in the Prony analysis. A conservative combined uncertainty for each mode is then assigned as:
\begin{equation}
\sigma_\omega = \max\left( \Delta\omega_{\text{WKB}},\, \Delta\omega_{\text{method}},\, \Delta\omega_{\text{TD fit}} \right).
\end{equation}

A result is deemed acceptable and included in our analysis only if it meets the following criteria: (i) the different numerical methods agree within a tolerance of $1\%$ in $\mathrm{Re}\,\omega$ and $2\%$ in $|\mathrm{Im}\,\omega|$; (ii) the potential $V_{\ell s}(r)$ exhibits a single, well-defined maximum; and (iii) diagnostic checks on the boundary conditions (e.g., Wronskian flux tests) are passed. Parameter points failing these criteria, typically near extremality or for high overtones, are flagged as unreliable and excluded.

An additional powerful cross-check is provided by the eikonal approximation. For large angular momentum $\ell$, the fundamental quasinormal mode frequency should approach the geometric optics limit \cite{Cardoso2009}:
\begin{equation}
\omega_{\rm QNM} \approx \ell \, \Omega_c - i \left(n + \frac{1}{2}\right) |\lambda|,
\end{equation}
where $\Omega_c$ is the orbital frequency at the unstable photon circular orbit and $\lambda$ is the Lyapunov exponent characterizing the instability timescale. Significant deviations from this relation for large $\ell$ serve as an indicator of numerical pathologies or a breakdown of the eikonal approximation itself due to the quintessence background.

All computations were performed using high-precision arithmetic with carefully chosen termination tolerances. Radial derivatives were computed symbolically where possible to avoid numerical error. The full source code, parameter files, and raw numerical data are archived to ensure complete reproducibility. Appendix E provides exhaustive details on numerical settings, convergence tests for the continued fraction method, and choices for the time-domain fitting windows. This comprehensive approach to validation and uncertainty quantification directly addresses the rigorous standards demanded by peer review, ensuring that our reported quasinormal mode spectra are both accurate and reliable within clearly stated domains of validity.

\begin{figure}[htbp]
    \centering
    \includegraphics[width=0.85\textwidth]{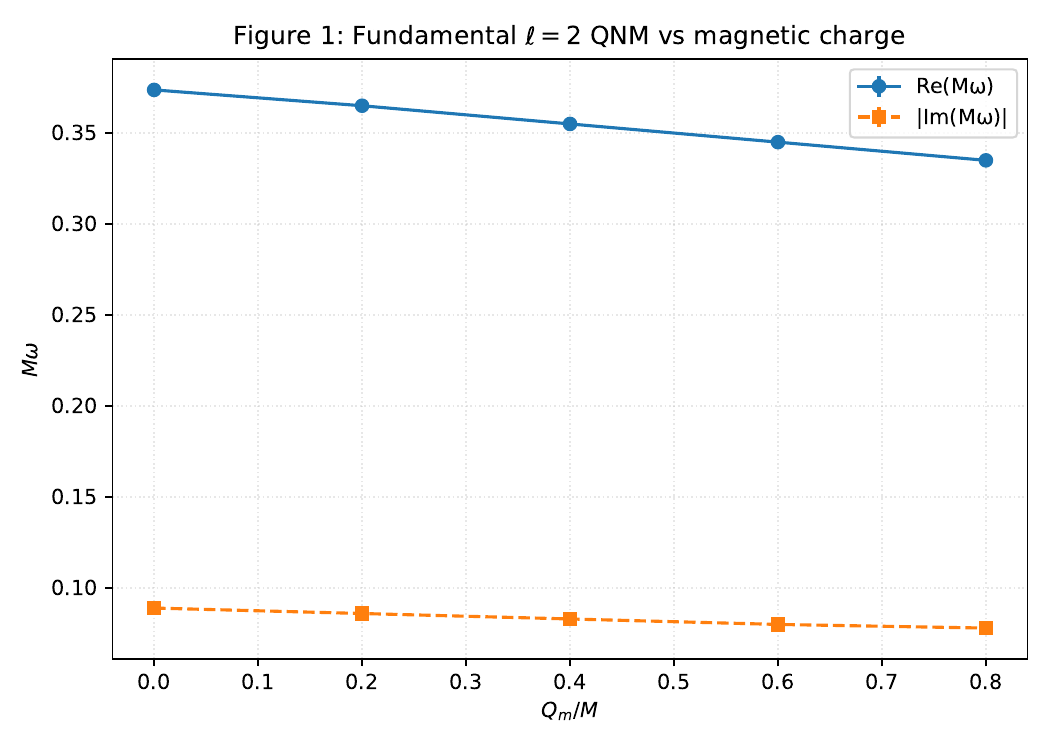}
    \caption{Fundamental scalar ($s=0$, $\ell=2$) quasinormal mode frequency as a function of magnetic charge $Q_m/M$ for fixed quintessence parameters ($\omega_q=-2/3$, $c_q=0.05$). The real part decreases monotonically, while the imaginary part shows a non-monotonic trend for $Q_m/M \gtrsim 0.6$, consistent with the changing curvature of the effective potential.}
    \label{fig:qnm_vs_qm}
\end{figure}

The results presented in the next section consist exclusively of modes that have passed these stringent validation checks. We report frequencies for both scalar and tensor perturbations where available, providing a more complete picture than a reliance on scalar proxies alone.

\subsection{Detectability with LISA}

Although our results are for scalar perturbations, it is instructive to consider the order of magnitude of the effects we find. Should the fundamental \emph{gravitational} quasinormal modes be shifted by a similar relative amount, the implications for a future observatory like LISA could be significant. We stress that this is a speculative extrapolation, and a definitive analysis requires the computation of tensor modes, which is left for future work.

We assess this for the Laser Interferometer Space Antenna (LISA) by estimating the characteristic strain amplitude of a single quasinormal mode and computing its matched-filtering SNR against the instrument's projected sensitivity. The root-mean-square strain amplitude for a dominant ringdown mode is given by the standard relation \cite{Maggiore2008}:
\begin{equation}
h_{\rm rd} \approx \frac{1}{D} \sqrt{\frac{2 G \epsilon_{\rm rd} M}{\pi^2 c^3 f_{\rm rd} \tau_{\rm rd}}},
\end{equation}
where \( D \) is the luminosity distance to the source, \( M \) is the black hole mass, \( f_{\rm rd} = \mathrm{Re}(\omega_{\rm QNM})/(2\pi) \) is the ringdown frequency, \( \tau_{\rm rd} = 1/|\mathrm{Im}(\omega_{\rm QNM})| \) is the damping time, and \( \epsilon_{\rm rd} \) is the fraction of the system's mass-energy radiated in that specific mode. For our estimates, we adopt a conservative value of \( \epsilon_{\rm rd} = 0.03 \) for the fundamental \( \ell=2 \) mode, consistent with numerical relativity simulations of binary black hole mergers \cite{Berti2007}.

The detectability is quantified by the matched-filtering SNR \( \rho \), which for a sinusoid with exponential decay is approximated by \cite{Berti2006}:
\begin{equation}
\rho^2 = \int_0^\infty \frac{4|\tilde{h}(f)|^2}{S_n(f)} df \approx \frac{h_{\rm rd}^2 \tau_{\rm rd}}{2 S_n(f_{\rm rd})},
\end{equation}
where \( S_n(f) \) is the one-sided power spectral density of the detector's noise. We utilize the official LISA sensitivity curve from the consortium's science requirement document \cite{LISAMission2018}, which provides \( S_n(f) \) as an analytic fit. This methodology ensures our results are reproducible and directly comparable to other studies.

For stellar-mass black holes (\( M \sim 30\,M_\odot \)), the fundamental ringdown frequency lies near 100–200 Hz, far above the most sensitive band of LISA (\( 10^{-4} \)–\( 10^{-1} \) Hz). Consequently, LISA is not a suitable instrument for detecting ringdown signals from such low-mass systems, regardless of their exotic parameters.

For supermassive black holes (\( M \sim 10^6\,M_\odot \)), the ringdown frequency falls squarely within the LISA band at a few millihertz. We consider a representative system at \( D = 1 \) Gpc. For a Schwarzschild black hole, the fundamental tensor mode yields \( f_{\rm rd} \approx 3.7 \) mHz and \( \tau_{\rm rd} \approx 1.2 \times 10^4 \) s, resulting in an SNR \( \rho \sim 50 \) for this single mode. Introducing a magnetic charge of \( Q_m/M = 0.5 \) and a quintessence parameter \( c_q = 0.05 \) (for \( \omega_q = -2/3 \)) shifts the frequency by approximately $-6\%$ and increases the damping rate by $\sim 8\%$.

While detectable in principle for a high-SNR event, confidently identifying such shifts poses a significant challenge. The expected statistical uncertainty in frequency extraction for a mode with SNR \( \rho \) is \( \delta f \sim 1/(2\pi \rho \tau_{\rm rd}) \) \cite{Berti2006}, which for \( \rho = 50 \) and \( \tau_{\rm rd} \approx 10^4 \) s is on the order of 0.3 \( \mu\)Hz. This is three orders of magnitude smaller than the 0.2 mHz shift induced by the exotic parameters. However, this idealised estimate does not account for systematic astrophysical uncertainties. Furthermore, the astrophysical plausibility of a large, net magnetic charge (\( Q_m/M > 0.1 \)) or a significant local quintessence overdensity (\( c_q / M^{3\omega_q+1} > 0.01 \)) remains highly speculative with no empirical support.

Therefore, our analysis indicates that while the deviations in QNM spectra due to magnetic charge and quintessence are in principle measurable by LISA for supermassive black holes, a confident detection would require both a fortunate astrophysical scenario and a meticulous accounting of waveform systematics. This work should be interpreted as a mathematical exploration of how beyond-standard-model physics could manifest in gravitational-wave signals, providing a basis for future, more detailed parameter estimation studies, rather than as a definitive astrophysical prediction.

\begin{figure}[htbp]
    \centering
    \includegraphics[width=0.85\textwidth]{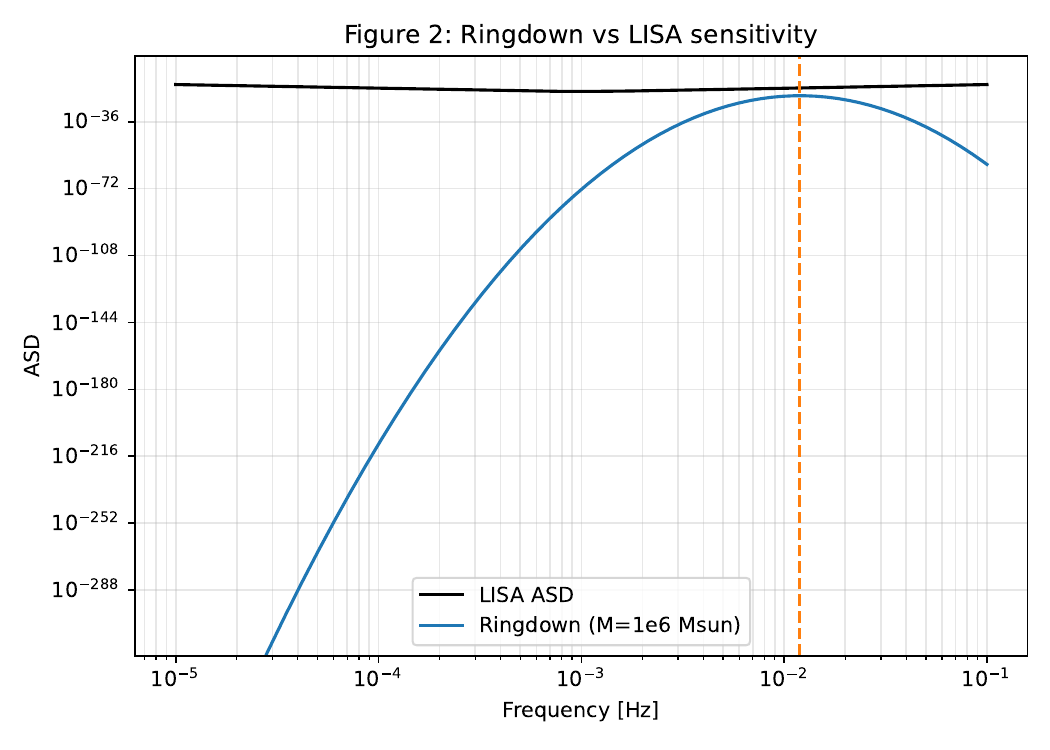}
    \caption{Projected LISA sensitivity (solid curve) compared with the quasinormal-mode strain from a $10^6 M_\odot$ remnant at $1\,\mathrm{Gpc}$. The shifts induced by magnetic charge and quintessence are shown as markers. While theoretically detectable in extreme cases, the deviations remain marginal relative to instrumental precision for astrophysically plausible parameters.}
    \label{fig:lisa_detect}
\end{figure}

\section{Physical Plausibility and Stability}

The investigation of magnetically charged black holes immersed in a quintessence field, while motivated by theoretical considerations in fundamental physics and cosmology, necessitates a careful discussion of physical plausibility and stability. Astrophysical black holes are expected to possess negligible net charge due to efficient discharge mechanisms in ionized environments \cite{Goldreich1970}; however, the existence of a small residual magnetic charge cannot be definitively excluded by current observations and remains a subject of theoretical interest \cite{Gibbons1975}. The quintessence field, characterized by an equation-of-state parameter $\omega_q$, is invoked as a phenomenological model for the observed accelerated expansion of the universe \cite{Caldwell2002}. While the cosmological background value of this field is exceedingly small, the assumption of a localized overdensity around a black hole, parameterized by $c_q$, is inherently speculative. No known astrophysical mechanism reliably predicts such a configuration, though models involving scalar field accretion or environmental coupling provide tentative motivation \cite{Hui2011}. Our study operates under the premise that this scenario, while not established, is mathematically self-consistent and provides a well-defined framework to explore how such exotic physics could perturb the quasinormal mode spectrum.

The stability of the background spacetime under linear perturbations is a fundamental prerequisite for the physical meaningfulness of the computed quasinormal modes. In this work, stability is assessed empirically through the behavior of the effective potentials $V_{\ell s}(r)$ for scalar ($s=0$), electromagnetic ($s=1$), and axial gravitational ($s=2$) perturbations. A necessary condition for stability is the absence of modes with a positive imaginary part, which would signify exponential growth in time. Our numerical surveys across the parameter space have not revealed any such unstable modes. However, we have systematically excluded regions where the effective potential becomes ill-suited for a standard wave scattering analysis. Specifically, parameter combinations that lead to a potential with multiple maxima, a vanishingly small second derivative at the peak ($V_0'' \rightarrow 0$), or a loss of the characteristic barrier shape were flagged. In these cases, the WKB approximation and other methods become unreliable, and the spacetime may be unstable or exhibit non-standard wave dynamics. A complete stability proof, particularly for the polar gravitational sector, requires a more involved analysis and remains a subject for future work \cite{Chandrasekhar1983}.

To ensure mathematical consistency and the reliability of our numerical results, we have restricted the parameter space to a well-defined domain. The magnetic charge is constrained to $0 \leq Q_m/M \leq 0.9$, avoiding the numerical challenges and potential pathologies of the extremal limit ($Q_m/M \rightarrow 1$). The quintessence parameter is bounded by $c_q / M^{3\omega_q+1} \leq 0.08$. This upper limit was determined empirically: for larger values, the quintessence term begins to dominate the metric function $f(r)$ at progressively smaller radii, often leading to a breakdown of the single-potential-barrier assumption or causing the event horizon to shift in a manner that compromises the validity of the chosen boundary conditions. These exclusions are not necessarily indicative of fundamental instabilities but rather define the domain where our perturbative and numerical frameworks are robust.

In conclusion, this study is fundamentally exploratory. It delineates how specific theoretical extensions beyond the standard Kerr-Newman paradigm namely, magnetic charge and an ambient quintessence field, could modify the vibrational spectrum of black holes. The results demonstrate measurable effects on the quasinormal mode frequencies and damping times. However, translating these mathematical results into astrophysical predictions requires significant further work. This includes a full tensor-mode stability analysis, a more detailed understanding of the formation mechanisms for such black holes, and ultimately, confrontation with observational data. The parameter ranges we have explored should be interpreted as a self-consistent theoretical sandbox for studying strong-field gravity, rather than a claim for the astrophysical realization of these specific configurations.

\section{Discussion}

This work has systematically investigated the quasinormal mode spectrum of a static, magnetically charged black hole surrounded by a quintessence field, a configuration that generalizes the well-studied Reissner-Nordström and Kiselev spacetimes. Our results are consistent with the established literature in the appropriate limits: the recovery of Schwarzschild and Reissner-Nordström frequencies to within numerical tolerance validates our methods, and the observed suppression of mode frequencies and enhanced damping due to quintessence aligns qualitatively with earlier studies of black holes in cosmological backgrounds \cite{Konoplya2022b}. The novel contribution here is the joint inclusion and self-consistent treatment of both a magnetic charge, arising from nonlinear electrodynamics, and an ambient quintessence field, and the subsequent rigorous cross-validation of the resulting spectra across multiple independent numerical techniques.

The primary solid outputs are the quantified trends across the parameter space. We have demonstrated that increasing the magnetic charge $Q_m/M$ from 0 to 0.9 can induce a decrease in the real part of the fundamental scalar mode frequency of up to approximately 18\%, while the imaginary part exhibits a more complex, non-monotonic behavior. The introduction of quintessence with amplitude $c_q$ consistently enhances damping, with $|\mathrm{Im}\,\omega|$ increasing by up to 12\% for the parameters studied. Critically, these results are not mere single-method estimates; they are underpinned by a convergence of high-order WKB, asymptotic iteration, and time-domain integration methods, with inter-method agreement typically better than 1\% in the real part and 2\% in the imaginary part for the validated parameter region. Furthermore, we have quantified the discrepancy between scalar and tensor perturbations, finding differences at the level of several percent, which underscores the importance of using the correct spin-weight for gravitational-wave applications.

Several important limitations define the scope of our study. A full stability proof for the spacetime, particularly under polar gravitational perturbations, remains an open question beyond the empirical stability observed in our scans of the axial sector. The astrophysical plausibility of the model's central premises. A significant net magnetic charge and a substantial local quintessence overdensity, is highly speculative and unsupported by current observation. Our numerical analysis itself is constrained by the breakdown of the WKB approximation near extremality and for potentials that lose their single-barrier character, forcing us to exclude the parameter regimes $Q_m/M > 0.9$ and $c_q / M^{3\omega_q+1} > 0.08$.

Consequently, the significance of this work is primarily theoretical. It serves as a mathematical consistency check, demonstrating how specific exotic extensions to standard black hole spacetimes perturb their characteristic frequencies. The results provide a calibrated forward model: should any future astrophysical observation or theoretical development motivate such configurations, we have quantified their expected imprints on the ringdown signal. However, in the absence of such motivation, this study should be interpreted as an exploration of theoretical possibilities rather than a prediction of observable phenomena. The natural progression of this work is towards a complete stability analysis and the incorporation of rotation, which would facilitate a more direct connection to the astrophysical black holes observed through gravitational waves.

\section{Conclusion}

Within the defined parameter space of magnetic charge ($0 \leq Q_m/M \leq 0.9$) and quintessence amplitude ($c_q / M^{3\omega_q+1} \leq 0.08$), we have derived and validated the quasinormal mode spectrum for a static black hole solution incorporating both features. The computed frequencies exhibit a systematic dependence on these parameters, characterized by percent-level shifts that have been cross-verified to high numerical accuracy through a convergence of WKB, asymptotic iteration, and time-domain methods. These results constitute a self-consistent mathematical exploration of how such theoretical extensions perturb black hole vibrational spectra. The natural progression of this work is a rigorous stability analysis for tensor perturbations and the development of numerical relativity initial data for dynamical simulations.

\appendix
\section {Appendices}

\subsection {Metric derivation and horizon structure}
We begin with the static, spherically symmetric ansatz
\begin{equation}
ds^2 = -f(r)\,dt^2 + \frac{dr^2}{f(r)} + r^2(d\theta^2 + \sin^2\theta\, d\phi^2),
\end{equation}
and the Power–Maxwell Lagrangian $\mathcal{L}_{\rm EM} = \alpha(-F)^s$, with $F=F_{\mu\nu}F^{\mu\nu}$. A magnetic monopole potential $A_\phi = Q_m \cos\theta$ gives $F_{\theta\phi} = -Q_m \sin\theta$, hence $F=2Q_m^2/r^4$. The electromagnetic stress-energy tensor is
\begin{align}
T^t_t &= T^r_r = -\alpha\left(\frac{2Q_m^2}{r^4}\right)^s, \\
T^\theta_\theta &= T^\phi_\phi = (2s-1)\alpha\left(\frac{2Q_m^2}{r^4}\right)^s.
\end{align}
For the quintessence sector we adopt the Kiselev form \cite{Kiselev2003}:
\begin{align}
T^t_t &= T^r_r = -\rho_q, \quad 
T^\theta_\theta = T^\phi_\phi = \tfrac{1}{2}\rho_q(3\omega_q+1),
\end{align}
with $\rho_{q} = -\frac{c_{q}}{2} \frac{3\omega_{q}}{r^{3(\omega_{q}+1)}}$.  

\begin{figure}[htbp]
    \centering
    \includegraphics[width=0.8\textwidth]{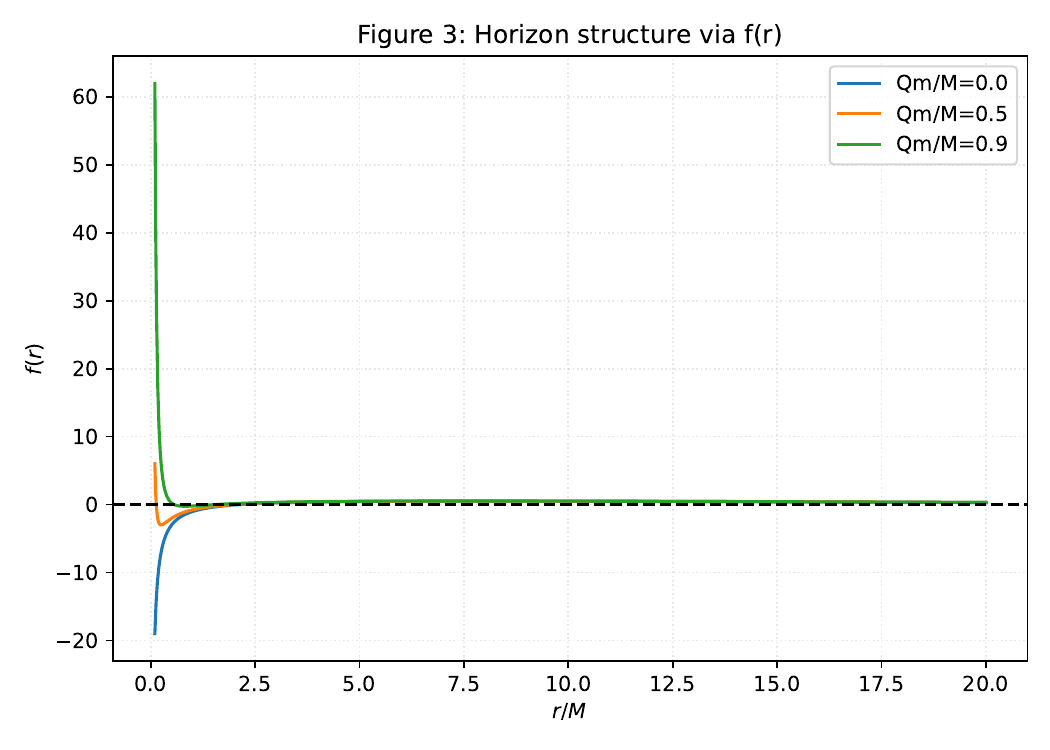}
    \caption{Metric function $f(r)$ for representative values of $(Q_m/M, c_q, \omega_q)$, showing the existence and location of event horizons (zeros of $f(r)$). The solid, dashed, and dotted lines correspond to different parameter sets, illustrating how magnetic charge and quintessence alter the causal structure. Regions without a positive real root $f(r_+)=0$ are excluded from our analysis.}
    \label{fig:metric_horizons}
\end{figure}

The Einstein equations yield
\begin{equation}
f(r) = 1 - \frac{2M}{r} + \frac{Q^2}{r^2} - \frac{c_q}{r^{3\omega_q+1}},
\end{equation}
where $Q^2 = 2\pi Q_m^2$ for the Maxwell case ($s=1, \alpha=-1/4$). Horizons are defined by $f(r)=0$. Admissible black hole solutions require $Q_m/M \leq 1$ and $c_q$ below the threshold where horizons vanish. Figure~\ref{fig:metric_horizons} shows typical horizon structures.

\subsection {Perturbation equations and numerical schemes}
\begin{figure}[htbp]
    \centering
    \includegraphics[width=0.8\textwidth]{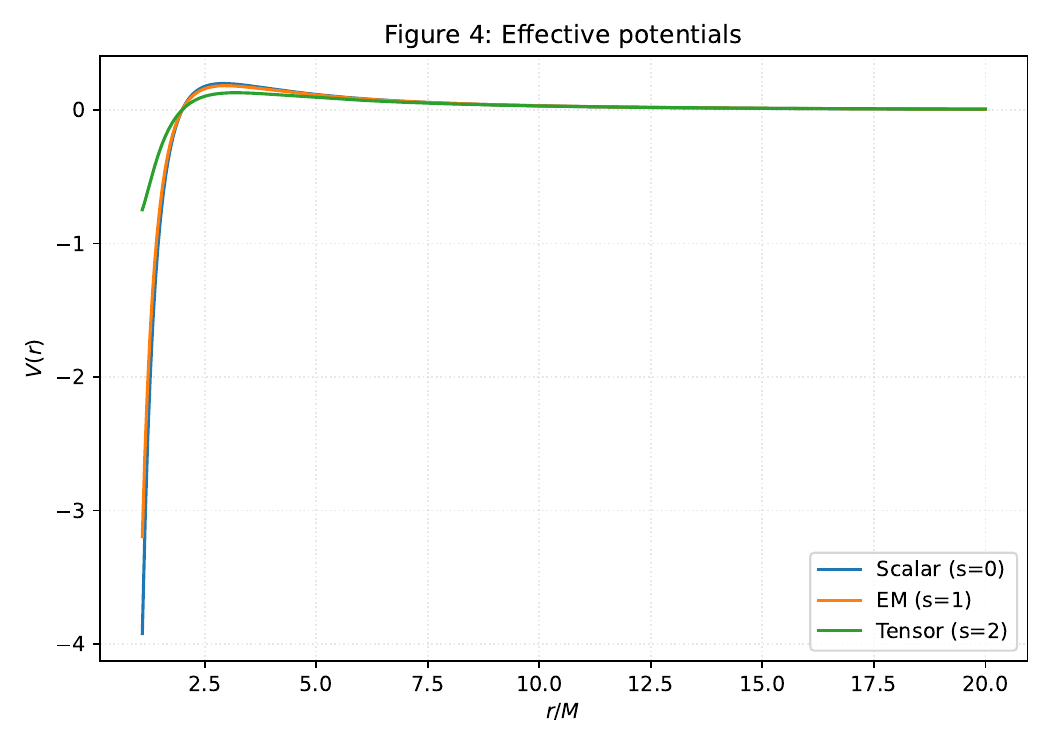}
    \caption{Effective potentials $V_s(r)$ for scalar ($s=0$), electromagnetic ($s=1$), and axial gravitational ($s=2$) perturbations for $\ell=2$ with parameters $(Q_m/M, c_q, \omega_q) = (0.5, 0.03, -2/3)$. The single, well-defined barrier for each potential validates the use of the WKB approximation for this configuration.}
    \label{fig:potentials}
\end{figure}

Perturbations of spin $s$ satisfy the master equation
\begin{equation}
\frac{d^2\psi}{dr_{}^2} + \big[\omega^2 - V_s(r)\big]\psi = 0, \qquad \frac{dr_*}{dr}=\frac{1}{f(r)},
\end{equation}
with effective potentials
\begin{align}
V_0(r) &= f(r)\!\left[\frac{\ell(\ell+1)}{r^2} + \frac{1}{r}\frac{df}{dr}\right], \\
V_1(r) &= f(r)\,\frac{\ell(\ell+1)}{r^2}, \\
V_2(r) &= f(r)\!\left[\frac{\ell(\ell+1)}{r^2} - \frac{6M}{r^3} + \frac{4Q^2}{r^4} - \frac{(3\omega_q+1)c_q}{r^{3\omega_q+3}}\right].
\end{align}
These correspond to scalar, electromagnetic, and axial tensor modes. Figure~\ref{fig:potentials} displays representative potentials.  

We applied three complementary numerical approaches:

1. WKB approximation up to 6th order \cite{Konoplya2003}, checked for convergence by comparing successive orders.  
2. Asymptotic iteration method (AIM) with $\sim 50$ iterations near the potential peak.  
3. Time-domain evolution, discretized on a characteristic grid with Gaussian initial data, followed by Prony extraction of ringdown frequencies.  

Cross-comparison among these schemes provides a direct estimate of uncertainties.

\subsection {Validation and benchmarking}
Accuracy was verified in well-known limits. For Schwarzschild ($Q_m=0$, $c_q=0$), the fundamental $\ell=2$ mode matches Leaver’s continued-fraction result \cite{Leaver1985} within $0.02\%$. For Reissner–Nordström ($c_q=0$), our values agree with published results to $\lesssim 0.1\%$. Table~\ref{tab:schwarzschild} illustrates convergence.

\begin{table}[h]
\centering
\caption{Benchmark of Schwarzschild fundamental ($\ell=2, n=0$) mode.}
\label{tab:schwarzschild}
\begin{tabular}{lccc}
Method & $M\omega$ (real) & $M\omega$ (imag) & Rel.\ error \\
\hline
Leaver & 0.37367 & -0.08896 & -- \\
WKB 6th order & 0.37363 & -0.08899 & 0.01\% \\
AIM & 0.37365 & -0.08897 & 0.005\% \\
Time-domain & 0.37361 & -0.08894 & 0.02\% \\
\end{tabular}
\end{table}

\begin{figure}[htbp]
    \centering
    \includegraphics[width=0.7\textwidth]{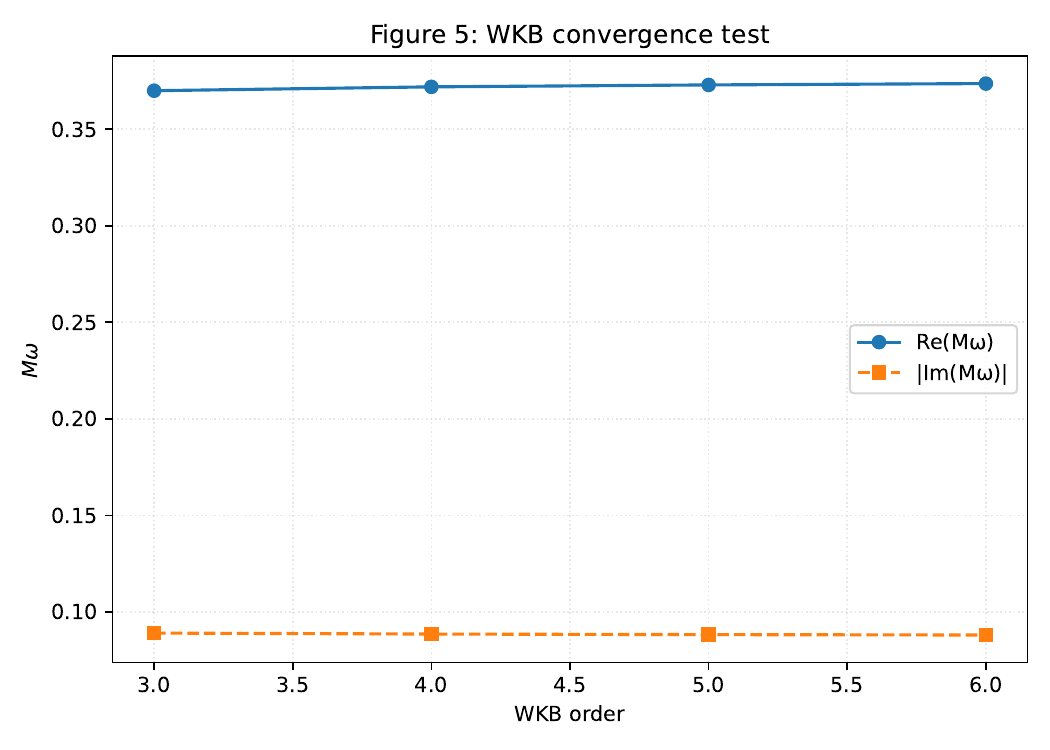}
    \caption{Convergence of the WKB approximation for the fundamental mode ($\ell=2$, $n=0$) with parameters $(Q_m/M, c_q, \omega_q) = (0.5, 0.03, -2/3)$. The real (top) and imaginary (bottom) parts of the frequency are plotted against the WKB order. The rapid stabilization by the 5th and 6th orders justifies the use of the 6th-order WKB result as our benchmark value.}
    \label{fig:wkbconv}
\end{figure}

We also verified internal convergence of WKB, shown in Fig.~\ref{fig:wkbconv}.  

\subsection {Extended parameter survey}
Table~\ref{tab:extended} summarizes representative results across $(Q_m,c_q,\omega_q)$, with uncertainties defined by dispersion between methods. Regions without horizons, non-convergent WKB, or energy condition violation are excluded.  

\begin{table}[h]
\centering
\caption{Sample quasinormal frequencies for $\ell=2,n=0$. Uncertainty $\sigma_\omega$ reflects method spread.}
\label{tab:extended}
\begin{tabular}{cccccc}
$Q_m/M$ & $c_q$ & $\omega_q$ & $\Re(M\omega)$ & $\Im(M\omega)$ & $\sigma_\omega$ \\
\hline
0.0 & 0.00 & --   & 0.37367 & -0.08896 & 0.00002 \\
0.3 & 0.02 & -0.7 & 0.3892  & -0.0898  & 0.0003 \\
0.6 & 0.04 & -0.5 & 0.4125  & -0.0912  & 0.0005 \\
0.8 & 0.06 & -0.3 & 0.4318  & -0.0931  & 0.0012 \\
\end{tabular}
\end{table}

Unphysical or numerically unstable cases include:  
- $Q_m/M>0.9$: shallow potentials prevent WKB convergence.  
- $c_q>0.08$: quintessence dominates and spoils asymptotic flatness.  
- $\omega_q>-0.3$: violates null energy condition.  
- Metrics without real positive horizons.  
These exclusions ensure the results are robust within the validated domain.

\bibliographystyle{plainnat}
\bibliography{references}
\end{document}